\documentclass[useAMS,usenatbib,usegraphicx]{mn2e}

\newif\ifAMStwofonts

\usepackage{lscape}

\newcommand{\kms}{km\,s$^{-1}$}

\newcommand{\around}{$\sim$}

\newcommand{\Msun}{M$_{\odot}$}

\newcommand{\MHIstar}{M$\rm ^{*}_{HI}$}

\newcommand{\Halpha}{H$\alpha$}

\newcommand{\Hbeta}{H$\beta$}

\newcommand{\Lya}{Ly$\alpha$}

\newcommand{\microJy}{$\rm \mu$Jy}

\newcommand{\arcminutes}{$^{\prime}$}

\newcommand{\arcseconds}{$^{\prime\prime}$}

\newcommand\aj{AJ}
\newcommand\araa{ARA\&A}
\newcommand\apj{ApJ}
\newcommand\apjl{ApJ}
\newcommand\aap{A\&A}
\newcommand\mnras{MNRAS}
\newcommand\nat{Nature}

\title{The HI content of star-forming galaxies at z = 0.24}

\author[Lah et al.]
{Philip Lah$^1$\thanks{E-mail: plah@mso.anu.edu.au}, 
Jayaram N. Chengalur$^2$, 
Frank H. Briggs$^{1,3}$, 
Matthew Colless$^4$,   
  \newauthor 
Roberto De Propris$^5$, 
Michael B. Pracy$^1$,
W. J. G. de Blok$^1$, 
Shinobu S. Fujita$^6$,
  \newauthor
Masaru Ajiki$^6$,
Yasuhiro Shioya$^7$,
Tohru Nagao$^8$, 
Takashi Murayama$^6$,
  \newauthor
Yoshiaki Taniguchi$^7$, 
Masafumi Yagi$^8$ 
and 
Sadanori Okamura$^9$ \\
\\
$^1$ Research School of Astronomy \& Astrophysics, The Australian National University, Weston Creek, ACT 2611, Australia \\ 
$^2$ National Centre for Radio Astrophysics, Post Bag 3, Ganeshkhind, Pune 411 007, India \\
$^3$ Australian Telescope National Facility, Epping, NSW 2111, Australia \\
$^4$ Anglo-Australian Observatory, PO Box 296, Epping, NSW 2111, Australia \\
$^5$ Cerro Tololo Inter-American Observatory, Casilla 603, La Serena, Chile \\
$^6$ Astronomical Institute, Graduate School of Science, Tohoku University, Aramaki, Aoba, Sendai 980-8578, Japan \\
$^7$ Physics Department, Graduate School of Science and Engineering, Ehime University, 2-5 Bunkyo-cho, Matsuyama 790-8577, Japan \\
$^8$ National Astronomical Observatory of Japan, 2-21-1 Osawa, Mitaka, Tokyo 181-8588, Japan \\
$^9$ Department of Astronomy and Research Center for the Early Universe, University of Tokyo, Bunkyo-ku, Tokyo 113-0033 \\
}  

\date{Accepted ...
      Received ...;
      in original form ...}

\pagerange{\pageref{firstpage}--\pageref{lastpage}}
\pubyear{2006}

\begin{document}

\maketitle

\label{firstpage}


\begin{abstract}

We use observations from the Giant Metrewave Radio Telescope (GMRT) to measure the atomic hydrogen gas content of star-forming galaxies at z~=~0.24 (i.e. a look-backtime of \around 3~Gyr).  The sample of galaxies studied were selected from \Halpha-emitting field galaxies detected in a narrow-band imaging survey with the Subaru Telescope.  The Anglo-Australian Telescope was used to obtain precise optical redshifts for these galaxies.  We then coadded the HI 21~cm emission signal for all the galaxies within the GMRT spectral line data cube.  From the coadded signal of 121 galaxies, we measure an average atomic hydrogen gas mass of $(2.26 \pm 0.90) \times 10^9$~\Msun .  We translate this HI signal into a cosmic density of neutral gas at z~=~0.24 of $\rm \Omega_{gas} = (0.91 \pm 0.42) \times 10^{-3}$.  This is the current highest redshift at which $\rm \Omega_{gas}$ has been constrained from 21~cm emission and our value is consistent with that estimated from damped \Lya\ systems around this redshift.  We also find that the correlations between the \Halpha\ luminosity and the radio continuum luminosity and between the star formation rate and the HI gas content in star-forming galaxies at z~=~0.24 are consistent with the correlations found at z~=~0.  These two results suggest that the star formation mechanisms in field galaxies \around 3~Gyr ago were not substantially different from the present, even though the star formation rate is 3~times higher.

\end{abstract}




\section{Introduction}

The cosmic star formation rate density is known to increase by an order of magnitude between the present time and z~\around~1 \citep{lilly96,madau96,hopkins04}.  However, the cosmic density of neutral gas is poorly observationally constrained over this redshift range.  Consequently, we are missing a key element in our understanding of galaxy evolution, that of the fuel supply for new stars.

In the local universe (z~\around~0) the quantity of atomic hydrogen gas in galaxies has been measured with good precision using HI 21~cm emission surveys \citep{briggs90,zwaan97,zwaan05}.  Observations of damped \Lya\ absorption in QSO spectra have proven very effective at measuring the density of atomic hydrogen at high redshifts, z~$> 1.5$ \citep{storrie00,prochaska05}.  However, it is difficult to constrain $\Omega_{\rm gas}$ from observations of damped \Lya\ absorbers at intermediate redshifts.  The difficulties are: (i)~that observations have to be done from space, since at these redshifts the \Lya\ line is not accessible by ground-based telescopes, and (ii)~that the average number of damped \Lya\ absorbers per unit redshift is so low that an accurate estimate of $\Omega_{\rm gas}$ requires impractical amounts of observing time. At lower redshifts (z~$< 1.5$), \citet{rao06} have made measurements of the neutral gas density using Hubble Space Telescope (HST) observations of clouds that were preselected to have strong MgII absorption.   However, this is a biased sample of objects compared to the random lines of sight probed by damped \Lya\ systems at high redshift.  In addition, the identification of the absorbing system in damped \Lya\ measurements is problematic, as the bright QSO light makes imaging or spectroscopy difficult of the fainter galaxies hosting the absorber. 

Using radio observations of HI 21~cm emission to constrain $\Omega_{\rm gas}$ at $\rm z > 0$ is difficult.  The problem is that the flux from most individual galaxies at these redshifts falls below the detection limit of current radio telescopes for reasonable observing times. Our method for bringing the measurement of the HI 21~cm emission within reach is to coadd the signal from multiple galaxies using their observed optical positions and redshifts.  This coadding technique for measuring atomic hydrogen gas has been used previously by \citet{zwaan00} in galaxy cluster Abell 2218 at $\rm z = 0.18$ and \citet{chengalur01} in galaxy cluster Abell 3128 at $\rm z = 0.06$.  We expand this technique to field galaxies with active star formation selected by their \Halpha\ emission.  With this technique, we can directly relate gas content, star formation rate and stellar mass (from continuum luminosities) in the same sample of objects.

The structure of this paper is as follows. Section~\ref{The_Data} details the optical and radio observations and data reduction. Section~\ref{Halpha_rc_correlation} comments on the relationship between the 1.4~GHz radio continuum and \Halpha~luminosities at z~=~0.24. Section~\ref{HI_results} details our main results, the HI content of star-forming galaxies at z~=~0.24.  Finally, Section~\ref{Conclusion} presents a brief summary and discussion. 

We adopt the consensus cosmological parameters of $\rm \Omega_{\lambda} = 0.7$, $\rm \Omega_{M} = 0.3$ and $\rm H_0 = 70 \ km \, s^{-1} \, Mpc^{-1}$ throughout the paper.  


\section{The Data} 

\label{The_Data}

\subsection{The Optical Data} 

\label{The_Optical_Data}

The optical galaxies we are examining come from a deep narrow-band imaging survey for \Halpha~emission galaxies at z~=~0.24 \citep{fujita03}.  This survey was performed using Suprime-Cam on the Subaru Telescope covering a field $34^\prime \times 27^\prime$ in size.  Deep observations were taken with broad passband filters B, R$_C$, I$_C$, z$^\prime$ and a narrow passband filter NB816 centred at 8150~\AA\ (corresponding to the \Halpha\ emission line at z~=~0.24).  After applying various selection criteria to remove contaminants,  \citeauthor{fujita03} found a total of 348 \Halpha-emitting galaxies with equivalent widths greater than 12~\AA\ (hereafter these galaxies will be referred to as the `Fujita galaxies'). From these galaxies, an extinction-corrected luminosity function for \Halpha\ emitting galaxies at z~=~0.24 was derived.  Using this luminosity function, \citeauthor{fujita03} computed a star formation rate density of $\rm 0.036^{+0.006}_{-0.012}\ M_{\odot}\ yr^{-1}\ Mpc^{-3}$. This is a factor of 3 higher than the local star formation rate density \citep{gallego95,condon02,gallego02}.  Their measured star formation rate density is in good agreement with other measurements made at and around z~=~0.24 \citep{hogg98,tresse98,haarsma00,pascual01,georgakakis03}.  This good agreement indicates that there is nothing particularly unusual about this field; it is not biased by containing an excess or deficiency of galaxies that have unusually high or unusually low star formation rates.

We observed the 348 Fujita galaxies with the 2dF and AAOmega spectrographs at the Anglo Australian Telescope to obtain precise spectroscopic redshifts.  Observations were carried out with 2dF on 6--10 March 2005, and an additional two hours of AAOmega service time obtained on 20~April 2006.  The data were reduced using the Anglo-Australian Observatory's {\bf 2dfdr} data reduction software\footnote{Anglo-Australian Observatory website \\ 
http://www.aao.gov.au/AAO/2df/aaomega/aaomega.html}.  
As the targets were selected for \Halpha\ emission, we confirmed their redshift primarily using the emission lines of [OII]$\lambda$3727, \Hbeta , [OIII]$\lambda$4959 and [OIII]$\lambda$5007. We obtained secure redshifts for 154 galaxies with $\rm 0.2150 < z < 0.2641$ and a maximum error in the redshift of 70~\kms .  The break down of the redshifts by \Halpha\ luminosity is: 47 redshifts for the 50 galaxies with L(\Halpha)~$>$~40.9~$\rm \log{\, erg \, s^{-1}}$, 48 redshifts for the 58 galaxies with 40.4~$<$~L(\Halpha)~$\le$~40.9, and 42 redshifts for the 240 galaxies with L(\Halpha)~$\le$~40.4.


\subsection{The Radio Data} 

 
\begin{figure}  

  \begin{center}  
  \leavevmode  
		
    \includegraphics[width=8cm]{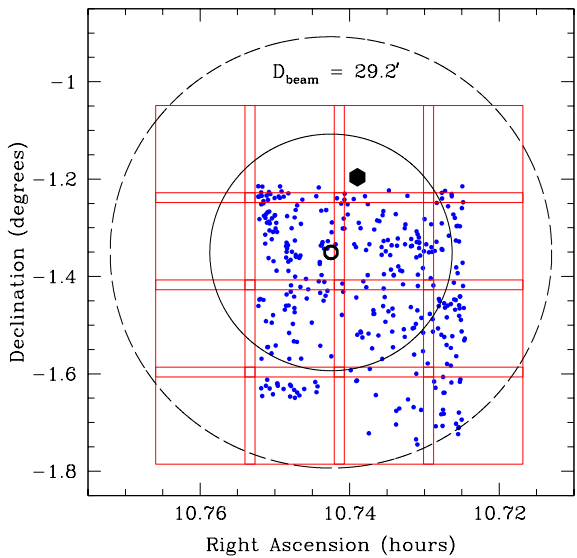}
  
   \end{center}

   \caption{This figure shows the positions of the optical galaxies as well as the pointing and beam size of the GMRT observations.  The 348 Fujita galaxies are the small points.  The large hexagonal point is the strong radio source J104420-011146 on which self calibration was performed.  The small, bold, open, central circle is the GMRT pointing of the observations. The unbroken circle is the primary beam diameter of 29.2' and the broken circle is the 10\% level of the beam.  The overlapping square grid shows the 16 facets used to tile the sky in the radio imaging.  The centre of this grid pattern was created with an offset from the GMRT pointing to ensure that all Fujita galaxies lie within the tiled region.}

   \label{SDF_positions}

\end{figure}


Radio observations of this field were carried out on 26--28 December 2003 and 7--9~\&~24--26 January 2004 using the Giant Metrewave Radio Telescope (GMRT) located near Pune in India.  A total of 80.5 hours of telescope time was used with \around 40 useful hours of on-source integration after the removal of slewing time, flux and phase calibrator scans, and the flagging of bad data. At most only 28 of the 30 GMRT antennas were providing useful data on any one day.  The total observing bandwidth of 32~MHz was split up into two 16~MHz-wide sidebands covering the frequency range from 1034~MHz to 1166~MHz (i.e. a redshift range $\rm 0.218 < z < 0.253$ for redshifted 21~cm HI~emission).  The pointing centre of the observations was R.A.~$\rm 10^{h}44^{m}33.0^{s}$ Dec.~$\rm -01^{\circ}21^{\prime}02^{\prime\prime}$ J2000 (see Figure \ref{SDF_positions}). There were two polarisations and 128 spectral channels per sideband, giving a channel spacing of 0.125~MHz (32.6~\kms).  Primary flux calibration was done using periodic observations of 3C286, which has a flux density at 1150~MHz of 16.24~Jy.  Phase calibration was done using scans on the VLA calibrator sources 0943-083 and 1130-148, for which our observations give flux densities of $3.137 \pm 0.011$~Jy and $5.163 \pm 0.044$~Jy respectively.

The data reduction was primarily done using AIPS and was partially automated using Perl scripts to interface with AIPS.  Each sideband of data were processed separately. The initial assessment of data quality and the preliminary calibration was done using the visibilities in a single channel. Flux and phase calibration was determined using the bandpass calibration task {\bf BPASS}, and the solution was interpolated to the data.  No normalisation was done before determining the solutions.  The regular phase calibrator scans throughout the observations allow us to correct for any time variability of the bandpass shape. Since, over the time span of the observations, there was no significant variation in the measured flux of the phase calibrators, the  measured fluxes for each phase calibrator were combined into a single average value.  In the final data reduction, this average value was used as the calibrated flux for the phase calibrators.  

After a period of testing and refinement, we adopted an imaging pixel size of 0.7\arcseconds\ and a robustness value of 0.  For primary beam correction, we assumed a Gaussian beam with FWHM  of 29.2\arcminutes\ at 1150~MHz\footnote{National Centre for Radio Astrophysics website \\ http://www.ncra.tifr.res.in/\around ngk/primarybeam/beam}.

The brightest radio continuum source in the field is J104420-011146. It lies \around 9.8\arcminutes\ from the phase centre and its measured integrated flux (at 1150~MHz) is $303.66 \pm 0.05 $~mJy. The integrated flux listed in the FIRST radio catalog (at 1.4~GHz) for this source is $276.83 \pm 0.15$~mJy. J104420-011146 is an order of magnitude brighter than the next brightest radio source in our field.  Self-calibration of the data was done using this source.

When making continuum images only channels 11-110 (out of the 128 in each sideband) were used. In order to avoid bandwidth smearing, the visibilities were averaged into a new data set consisting of ten channels, each of which was the average of 10 of the original channels.   This new ten channel u-v data file was turned into a single channel continuum image in the AIPS task {\bf IMAGR}, combining the channels using a frequency-dependent primary beam correction based on the antenna size of 45~m.  

 The self-calibration of GMRT data sometimes does not converge quickly, probably as a consequence of the GMRT hybrid configuration. Sources that lack coherence in the synthesis image due to phase errors tend to remain defocused.  To fix this problem, slightly extended sources were replaced in the first self-calibration loop with point sources with the same centroid and flux density as the original source. Further self-calibration loops were done using the clean components in the traditional manner. Each day's observation was initially self calibrated alone using 4 loops of self-calibration (2 loops of phase calibration and 2 amplitude and phase calibration loops).  At this point the bright continuum source J104420-011146 was subtracted from the u-v data, and the data were flagged to exclude any visibility residuals that exceeded a threshold set by the system noise statistics.  The bright continuum source was then added back into that day's data.

 
\begin{figure}  

  \begin{center}  
  \leavevmode  
		
    \includegraphics[width=8cm]{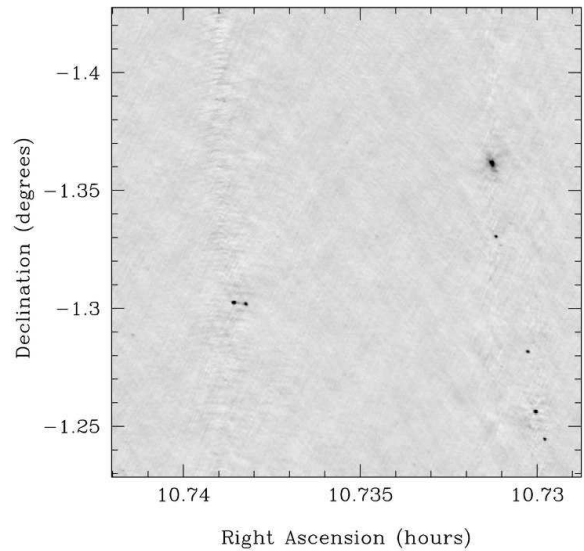}
  
   \end{center}

   \caption{This figure shows a grey scale image of one of the radio continuum facets (see the text for details).  This facet is just south of the facet with the strong continuum source J104420-011146 (see Figure~\ref{SDF_positions}).  The residuals from this bright source run north-south and can be seen on the left hand side of the image.  The RMS in the continuum images is 15~\microJy, but along this strip of residual structure, the RMS increases to \around 28~\microJy\ (this is the worst residual remaining in the any of the continuum facets).  The grey scale used in the image is a square root scale which varies from \around 2.5~mJy for black and \around -120~\microJy\ for white.  The brightest source in this image is in the top right hand corner and has extended structure.  It been identified as the nearly face-on spiral galaxy, UGC 05849 at z = 0.026.  The peak flux from the galaxy is 8.76 mJy/beam and the total flux density is $42.3 \pm 0.4$~mJy, which is spread over a diameter of \around 60\arcseconds .}

   \label{facet_10}

\end{figure}


After this detailed editing process, all the u-v data were combined and self-calibration was repeated on the entire combined data set to ensure consistency between the different days.  A large continuum image of the entire field was then made. 

The AIPS imaging routine assumes that the sky is flat (it ignores the vertical `w' term) which is appropriate only for small fields.  In order to reduce the distortion over our large field it was necessary to break the image up into 16~facets (see Figure \ref{SDF_positions}). Each of these facets was 11.95\arcminutes\ square (1024~pixels) and each overlapped by 1.2\arcminutes . The final total field size was 44.20\arcminutes\ on each side (i.e. large enough to include all the Fujita galaxies). Clean boxes were put around all the significant continuum sources during the imaging process.  In this initial continuum image, faint sidelobes from the brightest radio source J104420-011146 remained visible (see Figure~\ref{facet_10}). To reduce this, all the weaker continuum sources were subtracted from the u-v data, and the data were once again self-calibrated using clean components of the bright source alone.  This new model of the bright source was subtracted from the original u-v data, and the weak sources (along with the appropriate calibration for these sources) restored. The final continuum image of the field without the bright source was then made.  

The final image has a resolution of \around 2.9\arcseconds .  The RMS at the centre of the final continuum image is 15~\microJy . This is, to our knowledge, the most sensitive image achieved so far at the GMRT. The astrometry of the bright radio continuum sources in the data were checked against their positions in the VLA FIRST survey. A small offset of $(-0.79 \pm 0.30)$\arcseconds\ in right ascension and $(-1.42 \pm 0.28)$\arcseconds\ in declination was found.  This offset is believed to have been introduced to the data during the self-calibrating process though the exact origin of the offset is unclear.  After correcting for this, several objects could be identified with aligned optical and radio continuum components.  

The continuum emission was then subtracted from the original spectral line data using the AIPS task {\bf UVSUB}. While this removed most of the emission, small residuals of the brighter sources remained. To remove these final traces, a linear fit to the continuum across frequency was subtracted in the image plane using the AIPS task {\bf IMLIN}. This introduces a small bias, since any line emission from the galaxies would be included in the fit.  The correction for this bias is described in Section~\ref{HI_spectrum}.  In the final data cubes, the median RMS was 123~\microJy\ per channel in the lower sideband and 136~\microJy\ per channel in the upper sideband.  


\section{The Radio Continuum and \Halpha\ Luminosity Correlation}

\label{Halpha_rc_correlation}

 
\begin{figure}  

  \begin{center}  
  \leavevmode  
		
    \includegraphics[width=8cm]{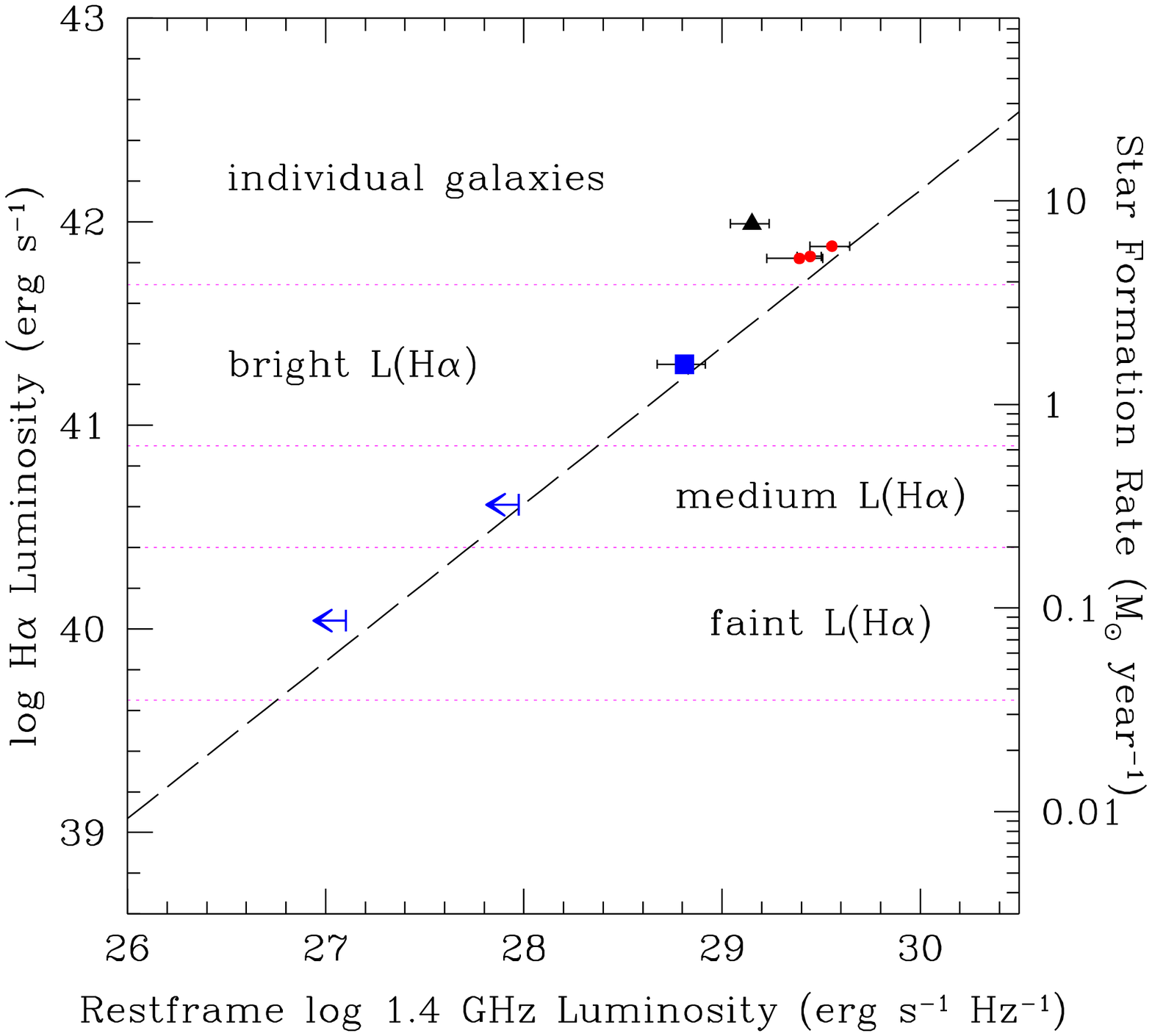}
  
   \end{center}

   \caption{
\Halpha\ emission line luminosity plotted against the restframe 1.4~GHz radio continuum luminosity for galaxies at z~=~0.24.   The line plotted through the data is the observed relationship found at z~=~0 (Figure 2 of \citealt{sullivan01}).  The 4 points in the `individual galaxies' region are the brightest \Halpha-emitting galaxies.  The triangle is a merging/interacting system of two galaxies.  The square in the `bright~L(\Halpha)' region is the combined signal from galaxies with bright \Halpha~emission.  The two radio continuum $2\sigma$ upper limits in the `medium~L(\Halpha)' and `faint~L(\Halpha)' regions are from the combined signal of subsamples of \Halpha-emitting galaxies.  The star formation rate shown on the right axis is derived using the \citet{kennicutt98} \Halpha\ to SFR relationship.
} 

   \label{Halpha_rc}

\end{figure}


Synchrotron radio emission is generated in areas of active star formation from relativistic electrons accelerated in supernova remnants.  Radio emission at 1.4~GHz generated by this process has been used as a measure of the star formation rate for galaxies, and it correlates well with other star formation indicators \citep{sullivan01}. In our radio continuum image, it is possible to measure this emission for some of the Fujita galaxies.  Since these galaxies are at a redshift of z~=~0.24 and we observed at 1150~MHz, the rest frequency of their radio continuum emission in our data is 1.4~GHz.  This means that we can directly compare the relationship in star forming galaxies between \Halpha\ line emission luminosity and 1.4~GHz radio continuum luminosity at z~=~0.24 and z~\around~0.  Our measurements are shown in Figure~\ref{Halpha_rc}.  Also shown in Figure~\ref{Halpha_rc} is the linear relationship found by \citet{sullivan01} from comparing the \Halpha\ emission against 1.4~GHz luminosity for 17 galaxies at z~\around~0. The uncertainties and the standard errors in the means for the \Halpha\ measurements in Figure~\ref{Halpha_rc} are all smaller than the size of the points.  

 
\begin{figure}  

  \begin{center}  
  \leavevmode  
		
  \includegraphics[width=8cm]{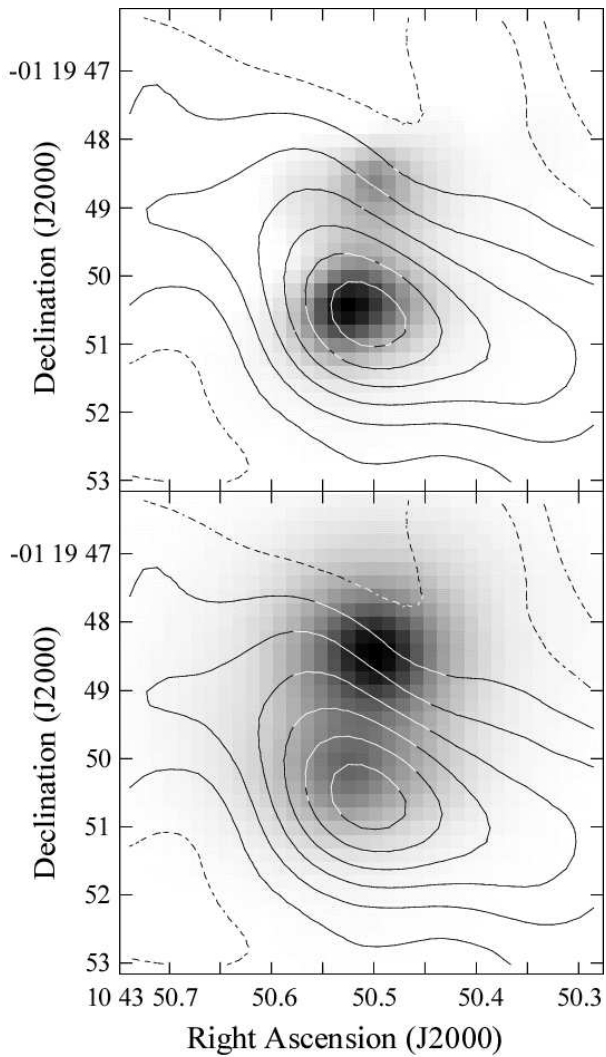}

   \end{center}

   \caption{The top figure shows the \Halpha~emission greyscale and radio continuum contours of the pair of merging/interacting galaxies mentioned in the text.  The same pair of galaxies are shown in the bottom figure with the same radio continuum contours and the `Iz continuum' (the `Iz continuum' is the combination of the Suprime-Cam I$\rm _C$ and z$^\prime$ images, the continuum level around the \Halpha\ line; see \citealt{fujita03} for details).  The radio continuum contour levels are -40, -20, 0, 20, 40, 60, 80, 100~\microJy .  The images are both 7\arcseconds\ on a side.} 

   \label{g97}

\end{figure}



The integral flux densities of the radio continuum sources were measured using the AIPS task {\bf JMFIT} which fits a 2D Gaussian to the source.  The radio continuum for the 4 brightest \Halpha-emitting galaxies each have individual measurable radio continuum fluxes (from \around 100 to \around 300~\microJy).  These are the points shown in the `individual galaxies' region of Figure~\ref{Halpha_rc}.  The three galaxies with circular points show reasonable agreement with the \citeauthor{sullivan01} line.  The triangle point is a merging/interacting system of two galaxies within 3\arcseconds\ (\around 11~kpc) of each other (see Figure \ref{g97}).  The galaxies were observed to be at the same redshift of z~=~0.2470 and have overlapping optical profiles.   Both galaxies show strong \Halpha\ emission but only the galaxy with brighter \Halpha\ emission shows any signs of radio emission.  Interestingly, this galaxy, although brighter in \Halpha , appears to be less massive than its companion (it is \around 5 times more luminous in \Halpha\ than its companion but is \around 0.6 times as luminous in the z' band filter). The reason why this galaxy's radio continuum luminosity lies away from the \citeauthor{sullivan01} line may have to do with it being a merging system.  

At lower \Halpha\ luminosities the radio continuum signal from multiple galaxies has been combined to improve the signal to noise ratio. Of the remaining 344 Fujita galaxies, 4 have been excluded as they have contamination by bright radio continuum sources within 10\arcseconds\ that do not appear to be connected to the galaxy.  Another 2 galaxies were excluded because they lie within regions of artifacts in the synthesis images.  Finally, we also exclude a galaxy with a very strong radio continuum flux ($2336 \pm 67$~\microJy, which is brighter than any other Fujita galaxy by an order of magnitude) but has a relatively low \Halpha\ flux ($\rm 39.82 \pm 0.10\ \log{\, erg \, s^{-1}}$).  The source of such a strong radio flux is more likely to be an active galactic nucleus than star formation.  Unfortunately we do not have confirmation of the redshift for this galaxy.

The galaxies have been divided up into three subsamples for the purpose of combining the radio continuum.  In these subsamples are: 45 bright galaxies with 40.9~$<$~L(\Halpha)~$\le$~41.69~$\rm \log{\, erg \, s^{-1}}$, 55 medium  galaxies with 40.4~$<$~L(\Halpha)~$\le$~40.9 and 236 faint galaxies with 39.65~$\le$~L(\Halpha)~$\le$~40.4.  Nearly all of the bright L(\Halpha) galaxies and most of the medium L(\Halpha) galaxies have optical redshifts, but the vast majority of the faint L(\Halpha) galaxies do not.  The radio continuum signal of the galaxies are combined using a weighted average, with the weight being the individual noise level for that galaxy.   The further a galaxy is from the centre of the GMRT observations the greater the beam correction to its flux density; this results in a higher noise level and thus a lower weight in the sum.

The square in Figure~\ref{Halpha_rc} is the average signal for the bright L(\Halpha) galaxies.  This value shows good agreement with the \citet{sullivan01} line.  For the medium L(\Halpha) and faint L(\Halpha) galaxies there was no detection of radio continuum flux, meaning that only upper limits can be determined.  The $2\sigma$ upper limits for these regions are plotted on Figure~\ref{Halpha_rc}.  When determining these upper limits, it was assumed that the galaxies were not resolved by the GMRT synthesised beam.  Both of these upper limits lie just to the left of the \citeauthor{sullivan01} line, suggesting that the linear relationship might not hold at low luminosities.  Indeed, \citeauthor{sullivan01} state that there is some evidence that the \Halpha --radio continuum relationship may not be perfectly linear for galaxies with low star formation rates.  They suggest that a fraction of the cosmic rays accelerated in supernovae remnants may escape from the lower mass galaxies.  This would reduce the radio continuum emission relative to the galaxy's \Halpha~luminosity.  This effect would explain the positions of the two lower limits without requiring a change in the star formation mechanism at z~=~0.24.

Overall, Figure~\ref{Halpha_rc} shows that at z~=~0.24 the relationship between radio continuum luminosity at 1.4~GHz and \Halpha~emission luminosity for galaxies is consistent with that found at z~=~0.  This suggests that there has been no significant change in the star formation mechanism over the last 3~Gyrs.


\section{HI Results}

\label{HI_results}

\subsection{The HI 21 cm Emission Signal}

\label{HI_spectrum}

 
\begin{figure}  

  \begin{center}  
  \leavevmode  
		
    \includegraphics[width=8cm]{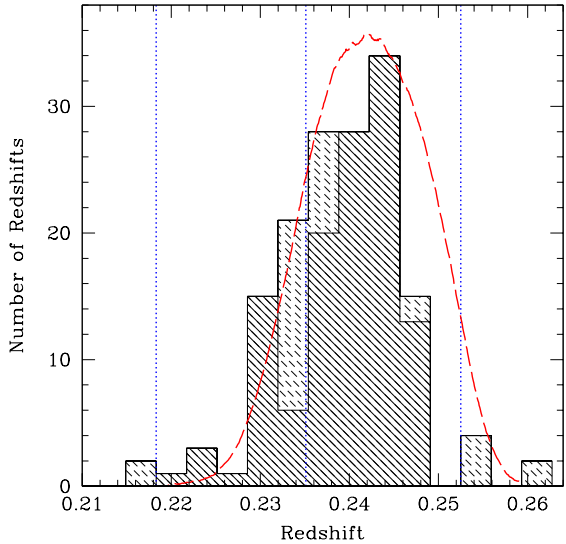}
  
   \end{center}

   \caption{The distribution of the measured redshifts for the Fujita galaxies.  There are 154 redshifts in total.  The dashed line is the shape of the Suprime-Cam narrow-band filter NB816 after converting the filter wavelength to the equivalent redshift of the \Halpha\ line.  The vertical dotted lines are the GMRT frequency limits converted to HI redshift; the central line is the boundary between the upper and lower sidebands of the GMRT data.  The histogram shaded with the unbroken line shows the distribution of the 121 redshifts usable for HI coadding.  The histogram shaded with the broken line shows the distribution of the 33 unusable redshifts (see text for details). } 

   \label{hist_z}

\end{figure}


To measure the coadded HI signal of the Fujita galaxies, we first use the optical redshifts to determine the expected frequency of the 21~cm HI emission.  Of the 154 optical galaxy redshifts obtained,  121 lie within the usable regions of the radio data cube (see Figure \ref{hist_z}).  The reasons the remaining 33 galaxies are unusable are either: (i)~their redshifted HI frequency lies outside the frequency range covered by the GMRT data, (ii)~their redshift lies close to the boundary of the two sidebands where the data quality is poor, or (iii)~their redshift lies within 5 channels of strong radio interference in the lower sideband (this strong radio interference is at \around 1137~MHz, covering \around 0.625 MHz).  For the purposes of defining a spectral window for coadding the HI signal, we allow for an average HI profile for the galaxies of the order of \around 300~\kms\ \citep{doyle05}. A velocity window width of 500~\kms\ should contain all the significant HI emission after factoring in the uncertainty in the optical redshifts (\around 70~\kms ).

In our data, the HI signals from individual galaxies are mostly below reasonable detection limits.  Using the HI mass function at z~\around~0 \citep{zwaan05}, we estimate that there is only a \around 15\% chance of a direct detection (5$\sigma$) of the HI flux from an individual galaxy within the volume probed by our GMRT observations.  There is probably more HI mass in galaxies at z~=~0.24 than at z~\around~0, making a direct detection more likely, however detecting significant numbers of galaxies is still unlikely.  For this reason we have coadded the HI signal from multiple galaxies to dramatically increase the signal to noise ratio of the measurement of the HI content of the sample.  

As galaxies are not individually detectable in our data, it is impossible to take into account any structure their HI gas may have when determining their HI flux density.  However, the peak specific intensity of an unresolved source in a radio synthesis image is equal to the flux density of that source.  This means, that if a galaxy is unresolved by the GMRT beam, we can take the value of the specific intensity at the optical position of the galaxy as a measure of the galaxy's total HI flux density.  

The GMRT synthesised beam has a FWHM \around 2.9\arcseconds\ which corresponds to \around 11~kpc  (1\arcseconds\ corresponds to 3.8~kpc at z~=~0.24).  All but the smallest of the galaxies in our sample are at least partially resolved by this beam size.  To solve this problem the radio data were smoothed to larger beam sizes of 5.3\arcseconds\ (\around 20~kpc) and 8.0\arcseconds\ (\around 30~kpc).  The larger galaxies should be unresolved in these larger beam size data. Gaussian smoothing in the image plane is equivalent to tapering (multiplying by a Gaussian) in the u-v plane. This effectively reduces the weight of the outer antennas in the image plane, and hence decreases the signal to noise ratio for unresolved sources.  However the measured HI flux increases for galaxies that are now unresolved in the smoothed data.  

A rough estimate of the HI size of each individual galaxy was made using a correlation between optical and HI size found by \citet{broeils1997}.  From their data, we derive a relationship between the diameter of the optical isophote corresponding to a surface brightness of 25~mag~arcsec$^{-2}$ in~B ($\rm D^{b,i}_{25}$) and the diameter within which half the HI mass of the galaxy is contained ($\rm D_{eff}$).  This relationship is:

\begin{equation}
  \rm \log(D_{eff}) = (0.978 \pm 0.035) \log(D^{b,i}_{25})\ + \ (0.041 \pm 0.046)
  \label{Deff}
\end{equation}

\noindent where $\rm D^{b,i}_{25}$ and $\rm D_{eff}$ are measured in kpc.  

The IRAF task {\bf ellipse} was used to measure the 25~mag~arcsec$^{-2}$ isophotal level in the Subaru $B$ images, taking into account the cosmological surface brightness dimming and a k-correction taken from \citet{norberg02}.  From this an estimate of the HI diameter is made.

 
\begin{figure}  

  \begin{center}  
  \leavevmode  
		
    \includegraphics[width=8cm]{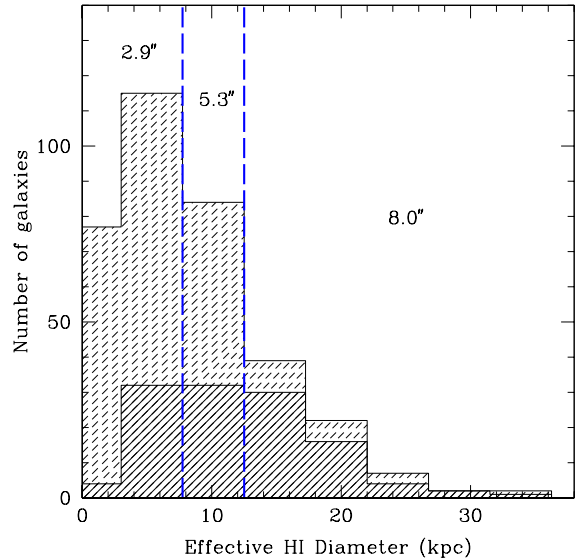}
  
   \end{center}

   \caption{This figure shows the histogram of the estimated effective HI diameter of the Fujita galaxies.  This was derived from the diameter of the galaxy at the optical B magnitude 25~mag~arcsec$^{-2}$ isophote and the correlation from \citet{broeils1997}.  The histogram shaded with the dashed line shows the distribution for all 348 Fujita galaxies; the histogram shaded with the solid line shows the distribution for galaxies with measured optical redshifts.  The vertical dashed lines show the boundaries used in sorting galaxies into the three different synthesised beam subsamples with beam FWHM of 2.9\arcseconds\ (11~kpc), 5.3\arcseconds\ (15~kpc) and 8.0\arcseconds\ (20~kpc). } 

   \label{hist_D_eff}

\end{figure}


If a galaxy has $\rm 2 \, D_{eff} < D_{mid}$, where $\rm D_{mid}$ is the midpoint between two beam smoothing levels, then it was assumed to be unresolved in the smaller beam data.  Figure~\ref{hist_D_eff} shows the distribution of the estimated effective HI diameter and the grouping of the galaxies into the three different beam sized radio data sets.
 
When making the final data cube, a linear fit to the spectrum through each sky pixel was subtracted to remove the residues left from the continuum sources (see discussion on {\bf IMLIN} previously).  Any spectral line features are included in the calculation of this linear fit across frequency.  Such features would create an over-estimation of the continuum, creating a bias in the measured HI spectrum.  To correct for this effect, each galaxy HI spectrum has a new linear fit made across all channels except those channels corresponding to a 500~\kms\ velocity width around the galaxy at its redshift (these should be the channels that contain any HI signal).  This new fit is then subtracted from the data, correcting for any bias created by {\bf IMLIN}.  This correction to the HI flux for the combined galaxy spectrum increases the measured flux by \around 25\%.  

When coadding the separate HI spectra, each measured flux value carries a statistical weight that is inversely proportional to the square of its uncertainty (the RMS noise level).  This noise value is calculated from the known RMS per frequency channel in the radio data cube, factoring in the increase in the noise from the beam correction for galaxies away from the centre of the GMRT beam.  As we are probing far from the GMRT beam centre, uncertainties in the beam shape could be a significant factor.  However as the coadded HI signal from individual galaxies is weighted by their noise, any errors in the assumed Gaussian shape in the outskirts of the beam are heavily attenuated by the weight factor. 

 
\begin{figure}  

  \begin{center}  
  \leavevmode  
		
    \includegraphics[width=8cm]{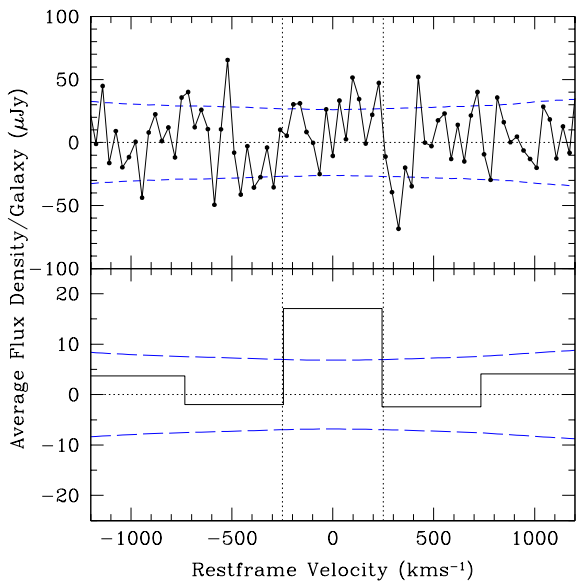}
  
   \end{center}

   \caption{The average HI galaxy spectrum created from coadding the signal of all 121 galaxies with known optical redshifts.  The top spectrum has no smoothing or binning and has a velocity step size of 32.6~\kms .  The bottom spectrum has been binned to \around 500~\kms .  This is the velocity width that the combined HI signal of the galaxies is expected to span.  For both spectra the $1\sigma$ error is shown as dashed lines above and below zero.}

   \label{HI_double}

\end{figure}


The weighted average HI spectrum from coadding all 121 galaxies can be seen in Figure~\ref{HI_double}.  To measure the error in the HI spectrum, a series of artificial galaxies with random positions and HI redshifts were used to create coadded spectra.  From many such artificial spectra a good estimate of the noise level in our final real spectrum can be determined.  As seen in Figure~\ref{HI_double}, the noise level increases with increasing velocity offset from the centre of the coadded spectrum.  This is because some galaxies lie at redshifts near the edges of the data cube and when adding the spectra of these galaxies to the total, there is no data for channels that lie off the edge of the data cube.  These channels with no data are given zero weight in the coadded sum resulting in a higher noise level at these velocities in the final coadded spectrum.  

In Figure~\ref{HI_double} the total HI flux density within a central velocity width of 500~\kms\ is $\rm 8.9 \pm 3.4 \ $mJy\,\kms , a signal to noise ratio of 2.6.  The level of this flux is not highly dependent on the choice of the velocity width; choosing various velocity widths from \around 400~to~\around 600~\kms\ gives total fluxes that are similar within the errors.  

HI emission flux can be converted to the mass of atomic hydrogen that produced the signal by the following relation:

\begin{equation}
  \rm M_{HI} = \frac{236}{( \ 1 + z \ )} 
  \left ( \frac{S_{v}}{ \ mJy \ }\right ) 
  \left ( \frac{d_{L}}{ \ Mpc \ } \right )^2  
  \left ( \frac{\Delta V}{ \ km \, s^{-1} }  \right ) 
\end{equation}

\noindent where $\rm S_{v}$ is the HI emission flux averaged across the velocity width $\rm \Delta V$ and $\rm d_L$ is the luminosity distance to the source.  This equation assumes that the cloud of atomic hydrogen gas has a spin temperature well above the cosmic background temperature, that collisional excitation is the dominant process, and that the cloud is optically thin \citep{wieringa92}.  No correction for HI self absorption has been made.   HI self-absorption may cause an underestimation of the HI flux by as much as 15\%.  However this value is extremely uncertain \citep{zwaan97}.  For our HI mass calculation we used $\rm \Delta V = 500$~\kms , mean z~=~0.235 and $\rm d_L = 1177$~Mpc.  This gives an average HI galaxy mass for all 121 galaxies of $(2.26 \pm 0.90) \times 10^9$~\Msun\ (Note: \MHIstar~=~$6.3 \times 10^9$~\Msun\ at z~=~0, \citealt{zwaan05}). 


\subsection{The Cosmic Neutral Gas Density, {\boldmath$\rm \Omega_{gas}$} }

\label{Omega_gas}


\begin{table*}  

\centering

\begin{centering}

\begin{tabular}[b]{|c|c|c|c|c|c|c|c|}  

\hline 

\ & 
{\bf Number} & 
{\bf L(\Halpha)} &
{\bf Average} & 
{\bf Average}  & 
{\bf Average}  & 
{\bf Average} &
{\bf Subsample}\\

{\bf Galaxy} & 
{\bf of} & 
{\bf Limits} &
{\bf L(\Halpha)} & 
{\bf SFR}  & 
{\bf HI Mass}  & 
{\bf Volume per} &
{\boldmath$\rm \Omega_{gas}$} \\

{\bf Subsample } & 
{\bf galaxies} & 
$\rm (\log{\ erg \ s^{-1}})$ &
$\rm (\log{\ erg \ s^{-1}})$ &
(\Msun\, yr$^{-1}$) &
$\rm (10^{9}\, M_{\odot})$  & 
{\bf galaxy}\ (Mpc$^3$) &
$(10^{-3})$\\
\hline

all galaxies &  
121 &
-- &
41.0 & 
0.79 &
$2.26 \pm 0.90$  & 
-- &
--  \\
\hline

bright L(\Halpha) &
42 &
L(\Halpha) $>$ 40.9 &  
41.4 & 
1.98 &
$4.1 \pm 2.2$ & 
$105 \pm 38$ &
$0.38 \pm 0.24$ \\  
\hline

medium L(\Halpha) &
42 &
40.4 $<$ L(\Halpha) $\le$ 40.9 &  
40.6 &  
0.31 &
$2.9 \pm 1.5$ &
\ $76 \pm 28$ &
$0.37 \pm 0.23$ \\ 
\hline

faint L(\Halpha) &
37 &
40.0 $\le$ L(\Halpha) $\le$ 40.4 &  
40.2 &  
0.13 &
$0.9 \pm 1.4$ &  
\ $53 \pm 27$ &
$0.16 \pm 0.26$ \\ 

\hline 

\end{tabular}
 
\caption{The table shows the properties of all the galaxies coadded together and the three \Halpha~luminosity subsamples.  For each subsample the average \Halpha~luminosity, average star formation rate (SFR) and average HI mass is listed.  Also listed is the average volume per galaxy which is the average volume one would need to probe in order find a single galaxy belonging to that luminosity subsample.   The reciprocal of this volume is the number density of such galaxies per Mpc$^3$.  The cosmological neutral gas density for each galaxy subsample is listed in the last column.  Combining these values gives the total cosmic neutral gas density.  The star formation rate listed is derived using the \citet{kennicutt98} \Halpha\ to SFR relationship.}

\label{HI_mass_table}  

\end{centering}
\end{table*}


The conversion of the HI mass of the galaxies to the cosmic density of neutral gas requires a measure of the number density of the galaxies.  This information is available from the \citet{fujita03} \Halpha\ luminosity function at z~=~0.24.   

The narrow-band filter used to select the Fujita galaxies is not square but has an almost Gaussian-like profile (see Figure \ref{hist_z}).  This means that galaxies with strong \Halpha\ emission can be detected in the wings of the filter but will have their measured \Halpha\ luminosity underestimated.  There is a strong luminosity bias in the spectroscopic redshifts we obtained due to this effect.  Bright galaxies that have \Halpha\ emission lines in the wings of the filter will be easier to obtain redshifts for than faint galaxies at the centre of the filter.  The distribution of redshifts appears to be slightly bias towards lower redshifts.  This small bias is probably caused by the [NII]6584 emission line contributing more to the narrow-band flux at the lower redshifts, giving rise to more detections.  Near z~\around~0.225, the emission from the [NII]6584 line and the [SII] doublet both come through simultaneously at opposite ends of the narrow-band filter.  This could explain the tail at low redshift seen in the redshift distribution.

\citet{fujita03} applied a statistical correction for the effect of the filter shape when creating their \Halpha~luminosity function.  This could be done because they have reasonably well-understood selection criteria for \Halpha~emission detection.  This is not the case for the Fujita galaxies for which we obtained redshifts.  Different galaxies were observed for different amounts of time, as it was not possible to configure the fibres in 2dF on all the galaxy targets simultaneously.  Some galaxies also had small errors in their astrometry that reduced the amount of light down the optic fibre obtained for that galaxy (these astrometric errors have since been corrected).  This uncertainty in the detection success rate for the galaxy redshifts is why we are using the \citeauthor{fujita03} \Halpha~luminosity function to determine the number density of the galaxies. 

For calculating the number density of the galaxies, the Fujita galaxies with redshifts have been divided into three subsamples based on their \Halpha\ luminosity.  In order to divide the galaxies accurately into subsamples, it is necessary to correct the \Halpha\ luminosity for the effect of the filter shape.  This is done in order to separate genuine faint galaxies from brighter galaxies seen through the wings of the narrow-band filter.  Without this correction, the faintest \Halpha\ emission galaxies are measured to have more atomic hydrogen gas than galaxies with stronger \Halpha\ emission.

An estimate of the necessary \Halpha\ luminosity correction can be made using the precise redshifts of the galaxies and the known shape of the narrow-band filter with wavelength\footnote{National Astronomical Observatory of Japan website \\ http://subarutelescope.org/Observing/Instruments/SCam/}.  The measured \Halpha\ luminosity is adjusted by the relative decrease in the filter transmission at the redshifted wavelength of the \Halpha\ line compared to the filter centre\footnote{This \Halpha\ correction was not used in the comparison of the \Halpha\ and radio continuum observations in Section \ref{Halpha_rc_correlation}.  Corrections cannot be done for all the galaxies included in that analysis as not all have redshifts.  The correction is only necessary in the work above because of the bias in the galaxies for which we obtained spectroscopic redshifts.  The much larger sample of faint galaxies in Section \ref{Halpha_rc_correlation}, most of which do not have redshifts, is dominated by genuine faint galaxies rather than bright galaxies shining through the edge of the narrow-band filter.  It is these bright galaxies that need the correction.  The number of these bright galaxies misclassified as faint galaxies should be small compared to the large number of actual faint galaxies in that analysis, so that any correction should not be statistically significant.}.   

Using the corrected \Halpha\ luminosities the galaxies are separated into three subsamples: bright, medium and faint.  The details for each of the \Halpha\ subsamples can be seen in Table~\ref{HI_mass_table}.  The average HI mass for the galaxies in each subsample is measured as described in Section~\ref{HI_spectrum}. 

An estimate of the average number density for galaxies belonging to each subsample is determined using the \citet{fujita03} luminosity function.  This is done by integrating the \Halpha\ luminosity function over the \Halpha\ limits for each subsample to obtain the number density of such galaxies.  The galaxy number density for each subsample is expressed in Table~\ref{HI_mass_table} as the average volume one would need to probe in order to find a single galaxy belonging to that \Halpha\ luminosity subsample (the reciprocal of the number density).  The errors in the calculated volumes are derived by considering the effect of the errors in the \citeauthor{fujita03} Schechter luminosity function parameters.

The cosmic density of HI gas for each subsample can be calculated using:   

\begin{equation}
  \rm \rho\,_{HI} = \frac{ \overline{M}_{HI} }{ \overline{V}_{gal} }
\end{equation}

\noindent where $\rm \rho\,_{HI}$ is the cosmic density of HI gas, $\rm \overline{M}_{HI}$ is the average HI mass per galaxy and $\rm \overline{V}_{gal}$ is the average volume per galaxy.  To compare with other values in the literature, it is necessary to convert this value of the HI gas density to the density of neutral gas.  This is done by adding a correction for the neutral helium content, which is assumed to be 24\% by mass of the total neutral gas. We convert this neutral gas density to units of the total cosmic mass density using:

\begin{equation}
  \rm  \Omega_{gas} = \frac{8 \, \pi \, G \, \rho_{gas}}{3 \, H_0^2}
\end{equation}

 
\begin{figure*}  

  \begin{center}  
  \leavevmode  
		
    \includegraphics[width=8cm]{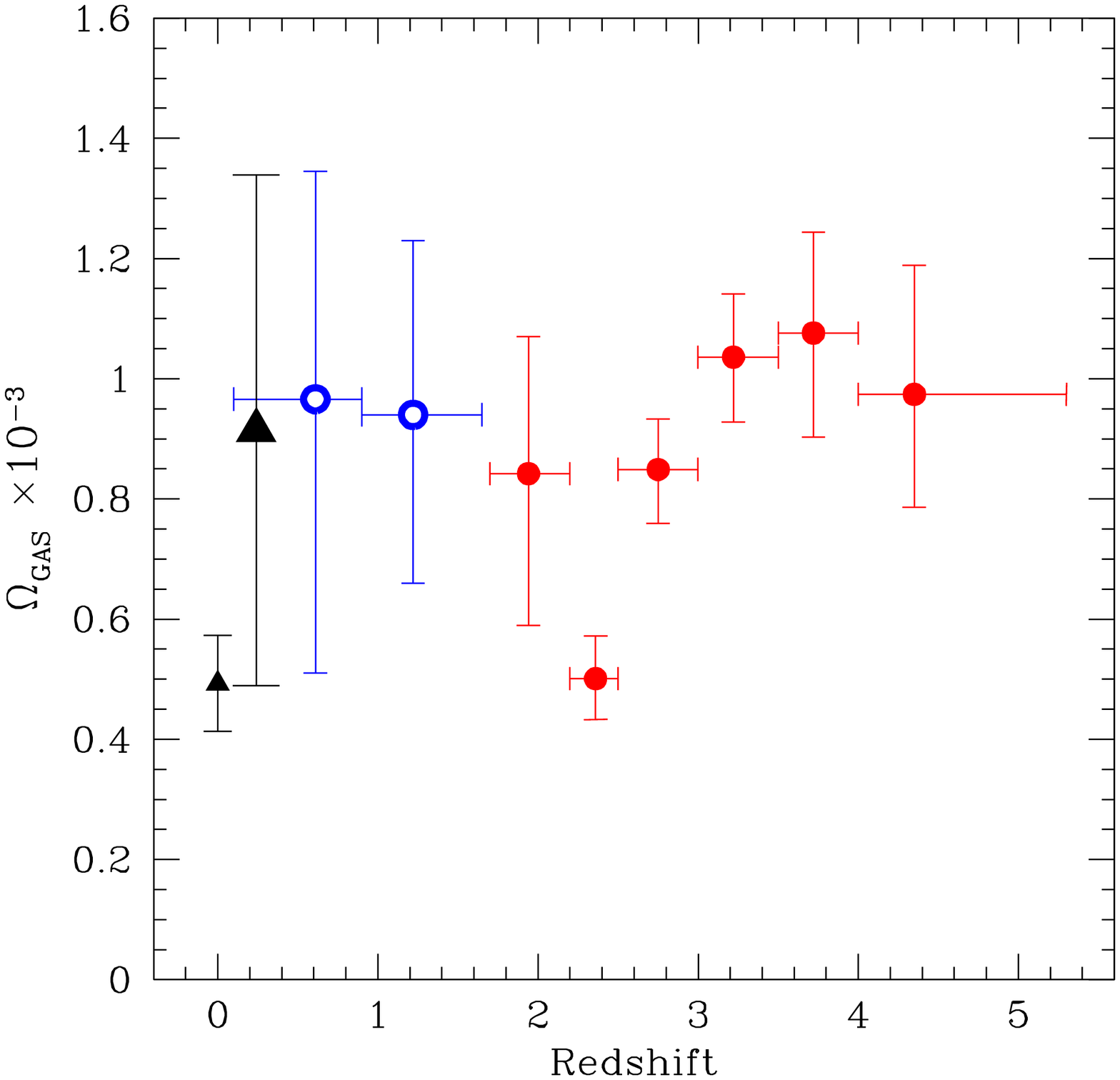}
    \includegraphics[width=8cm]{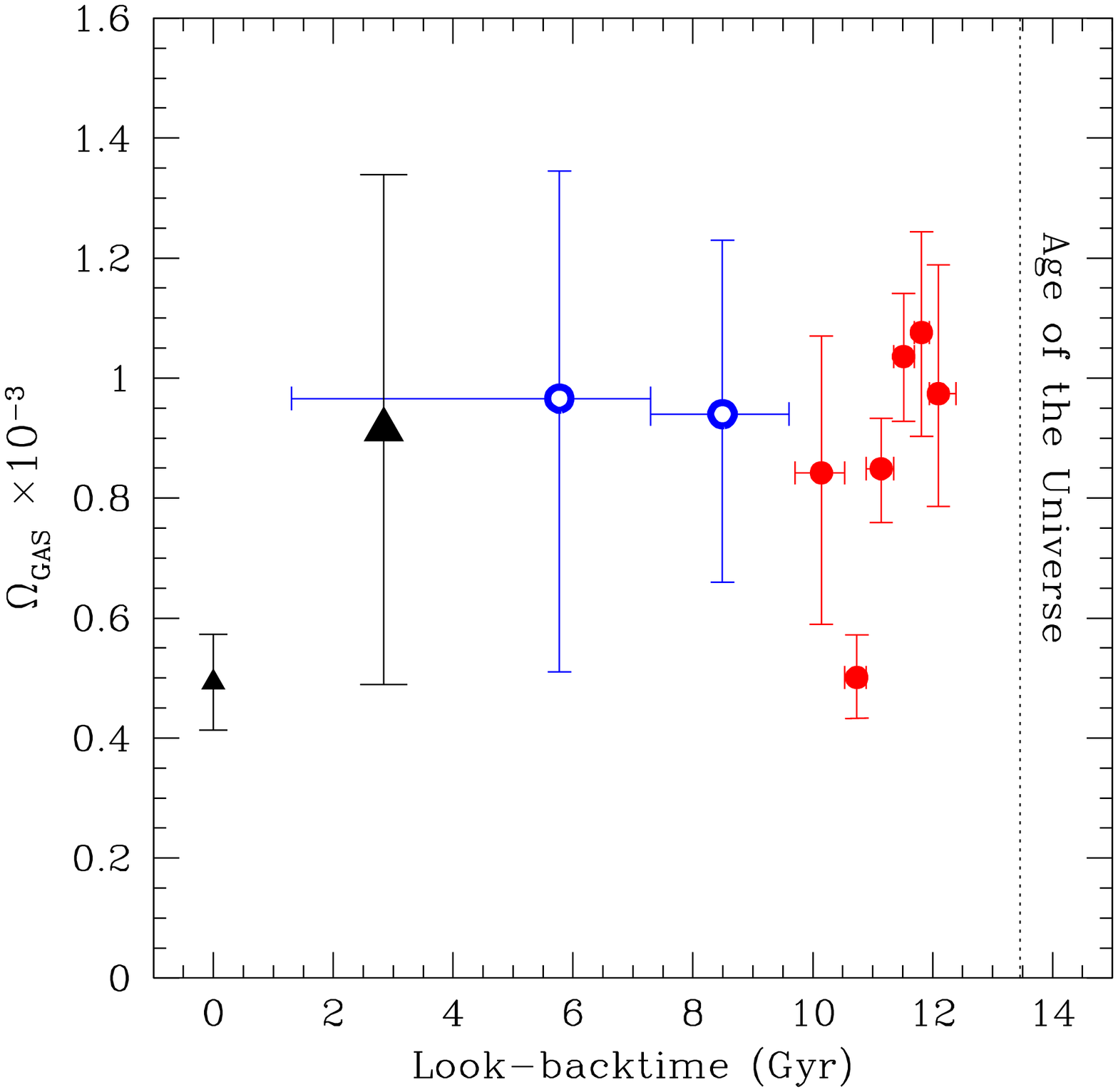}

   \end{center}

   \caption{The neutral gas density of the universe as a function of redshift (on the left) and look back time (on the right).  The small triangle at z~=~0 is the HIPASS 21~cm emission measurement from \citet{zwaan05}.  The filled circles are damped \Lya\ measurements from \citet{prochaska05}.  The open circles are damped \Lya\ measurements from \citet{rao06} using HST.    The large triangle at z~=~0.24 is our HI 21~cm emission measurement made using the GMRT.  All results have been corrected to the same cosmology and include an adjustment for the neutral helium content.}

   \label{neutral_gas_redshift}

\end{figure*}


\noindent After combining the neutral gas density values for each subsample we obtain a total cosmic density of neutral gas in star forming galaxies at z~=~0.24 of $\rm \Omega_{gas}~=~(0.91~\pm~0.42)~\times~10^{-3}$.  This value is shown in comparison to other literature values in Figure \ref{neutral_gas_redshift}.  

Figure \ref{neutral_gas_redshift} shows the neutral gas density of the universe as a function of both redshift and look-backtime.  There is a large uncertainty in the cosmic neutral gas density in the redshift range z~=~0.1--1.5, which corresponds to two thirds of the age of the universe.  Our measured value of $\rm \Omega_{gas}$ is comparable to that from previous damped \Lya\ measurements at intermediate redshifts \citep{rao06}.  While the trend of the measurements is to indicate a sharp rise in HI content with redshift, both studies suffer from large statistical uncertainty.  However, our new measurement has the advantage that it applies to a narrow range of redshifts z~\around~0.24.  

Strictly speaking, our measured cosmic neutral gas density is only a lower limit, as it includes only the gas from galaxies that show star formation.  The star formation--HI mass correlation found in low-redshift HI-selected galaxies \citep{doyle06}, shows that galaxies with little or no star formation contain no significant quantities of neutral gas.  This relationship appears to hold at z~=~0.24  (see Section \ref{SFR_gas}).  Since our sample includes galaxies with \Halpha\ luminosities at \around 1\% of L${_*}$  no significant reservoirs of HI gas should be missed.  

In this work no attempt has been made to separate the \Halpha~emitting galaxies into those that are genuine active star forming galaxies from those that are actually active galactic nuclei (AGN).  When calculating their star-formation density for this field at z~=~0.24, \citet{fujita03} included a 15\% correction for the luminosity contribution of AGN.  (This correction came from \citealt{pascual01}).  No similar correction exists for the atomic hydrogen gas content of AGN. However, the ratio of HI mass to \Halpha\ luminosity for AGN is likely to be lower than for star forming galaxies, as AGN preferentially reside in early-type galaxies \citep{schade01}. Including AGN in our sample may cause a small underestimation in the density of neutral gas for star forming galaxies, however it will bring our result closer to the true total cosmic neutral gas density of all galaxies by including another galaxy type in the measurement.  

The volume probed by \citet{fujita03} for their \Halpha\ luminoisty function (and used here for the calculation of the number density of galaxies) is \around 4000 Mpc and is a rectangular strip with a depth \around 110 Mpc. The cosmic variance in the number of galaxies in a field of this volume and shape at z~=~0.24 is approximately 40\%.  This estimate is based on a calculation using the \citet{eisenstein98} fitting function for the matter power spectrum functon and is confirmed by an examination of the galaxy distribution in the Millennium Simulation at z~=~0.24 \citep{springel05,croton06}. Treating this as a random error (rather than a systematic offset), the final error in $\rm \Omega_{gas}$ would increase by \around 15\%.  However, the star formation rate measured for this field places it close to the mean, or perhaps slightly higher, compared to other values in the literature (see Section \ref{The_Optical_Data}).  This suggests that this field has close to the average number of galaxies despite the large possible cosmic variation, and that the measured neutral gas density measured in this field is close to, or slightly higher than, the true cosmic gas density, $\rm \Omega_{gas}$. 


\subsection{The Star Formation Rate--HI Mass Galaxy Correlation}

\label{SFR_gas}

\begin{figure}  

  \begin{center}  
  \leavevmode  
		
    \includegraphics[width=8cm]{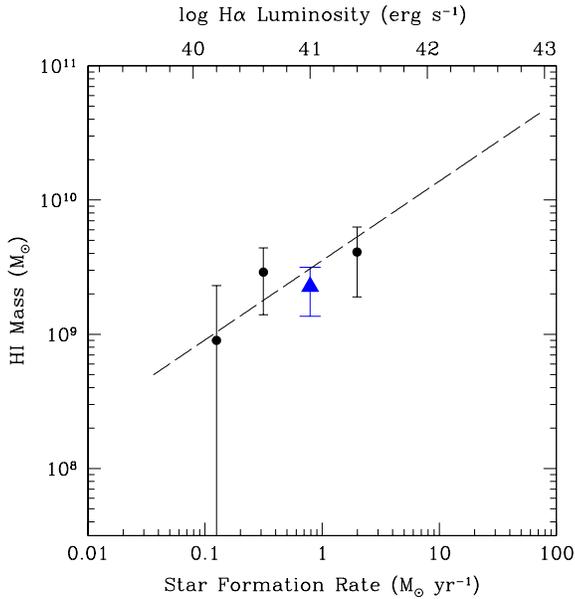}
  
   \end{center}

   \caption{
The average galaxy atomic hydrogen gas mass plotted against the average galaxy star formation rate.  The circular points are the average values for the bright, medium and faint L(\Halpha) subsamples  (see Table \ref{HI_mass_table}).  The triangle is the value for the average of all galaxies.  The line is from \citet{doyle06} for z~\around~0 galaxies.
} 

   \label{SFR_HI}

\end{figure}

In the local universe there is a known correlation between the star formation rate in a galaxy and the mass of HI in that galaxy.  For a recent study using HI-selected galaxies from HIPASS, see \citet{doyle06}.  In Table~\ref{HI_mass_table} we have listed the average HI galaxy mass for the different \Halpha~luminosity subsamples.   For each subsample, one can derive the average star formation rate per galaxy from the average \Halpha\ luminosity and the SFR conversion of \citet{kennicutt98}.  Our comparison of star formation rate and galaxy HI mass is shown in Figure~\ref{SFR_HI}.  Plotted on this figure is the linear relationship from Figure 3 of \citet{doyle06}, where they compared individual galaxies' HI masses from HIPASS to their star formation rate derived from {\it IRAS} infrared data.  The linear fit, taken directly from their figure, is:

\begin{equation}
\rm \log(HI\ Mass) = 0.59 \,\log(SFR) \, + \, 9.55
\label{HI_SFR}
\end{equation}

\noindent From Figure~\ref{SFR_HI}, it is clear that the values found at z~=~0.24 are consistent with the correlation found at z~=~0.  This suggests that there has been no significant change in the star formation mechanisms in field galaxies 3~Gyrs ago; that the same amount of atomic hydrogen gas in a galaxy gives the same measurable star formation rate. 

At z~$>$~1 there is a factor \around 10 increase in the star formation density compare to z~=~0 \citep{hopkins04}.  If the star formation--HI mass correlation as stated in equation~\ref{HI_SFR} were to hold true at these high redshifts, a factor of \around 4 increase in neutral gas density would be required.  However, the damped \Lya\ measurements of \citet{prochaska05} only indicate that the neutral gas density rises by factor of \around 2.  This implies that star formation rate--HI mass relationship must have been different at these higher redshifts, indicating a change in the mechanism of star formation.  The mechanism for converting gas to stars must have been a factor of \around 2 more efficient at high redshifts.  

 
\section{Conclusion}

\label{Conclusion}

We have demonstrated that is possible to measure, in reasonable observing times, the average amount of atomic hydrogen gas in galaxies from their HI 21~cm emission at intermediate redshifts.  This is done by coadding the HI signal from multiple galaxies with known positions and optical redshifts.  Our measurement of HI 21~cm emission from galaxies corresponds to a cosmic density of neutral gas that is consistent with previous measurements of the neutral gas density in damped \Lya\ measurements at a similar redshift.  

The relationship between \Halpha\ luminosity and the restframe 1.4~GHz radio continuum emission in star-forming galaxies at z~=~0.24 has been shown to be consistent with the correlation found at z~=~0, as is the the relationship between galaxy star formation rate and galaxy HI mass.  These two results suggest that the process of star formation in field galaxies is not significantly different 3~Gyr ago from the present day. 

However, the significantly higher global star formation rate density found at still higher redshifts does not seem to continue this trend.  At z~$> 1$, the damped \Lya\ measurements of the neutral gas density and the star formation rate density differ by a factor of two from what one would expect from the star formation--HI mass correlation found at z~=~0.  This implies an enhanced efficiency of star formation for a given neutral gas density at high redshifts.


\section*{Acknowledgments}

We thank the staff of the GMRT who have made these observations possible. The GMRT is run by the National Centre for Radio Astrophysics of the Tata Institute of Fundamental Research. We are grateful to the staff of the Anglo Australian Observatory staff for their assistance.  This work is based in part on data collected at the Subaru Telescope, which is operated by the National Astronomical Observatory of Japan.  We are indebted to Ayesha Begum, Agris Kalnajs, Nissim Kanekar, Brian Schmidt and Eduard Westra for their valuable help.



\label{lastpage}

\end{document}